\documentstyle[aps,amstex]{revtex}
\input epsf
\begin{document}

\title{On the topological classification of binary trees using the
Horton-Strahler index} 

\author{Zolt\'an Toroczkai}
\address{
Theoretical Division and Center for Nonlinear Studies,
Los Alamos National Laboratory, Los Alamos, New Mexico 87545} 

\date{\today}
\maketitle

\begin{abstract}
The Horton-Strahler (HS) index
$r=\max{(i,j)}+\delta_{i,j}$ has been shown to be relevant to a
number of physical (such at diffusion limited aggregation) geological
(river networks), biological (pulmonary arteries, blood vessels, various
species of trees) and computational (use of registers) applications.
Here we revisit the  enumeration problem of the HS index on
the rooted, unlabeled, plane binary set of trees, and
enumerate the same index on the ambilateral set of rooted,
plane binary set of trees of $n$ leaves. The ambilateral set
is a set of trees whose elements cannot be obtained from
each other via an arbitrary number of reflections with respect
to vertical axes passing through any of the nodes on the tree. For the
unlabeled set we give an alternate derivation to the existing exact
solution.
Extending this technique for the ambilateral set, which is described
by an infinite series of non-linear functional equations, we are able
to give a double-exponentially converging approximant to the
generating functions in a neighborhood of their convergence
circle, and derive an explicit asymptotic form for the number of
such trees.
\end{abstract}

\pacs{PACS numbers:} 
\vspace*{-0.8cm}

\section{Introduction}

Trees are ubiqutous structures which appear naturally in a
large number of
physical, chemical, biological and social phenomena, such as river
networks, diffusion limited aggregation, pulmonary arteries,
blood vessels and tree species, social organizations, decision
structures, etc. They also play an important role in
computer science (use of registers and computer languages), in graph
theory, and in various methods of statistical physics such as
cluster expansions and renormalization group.

In spite of their apparent structural simplicity, and the large body
of scientific work on trees (a sample of which is found in \cite{Cayley}-\cite{Flajolet},
\cite{bkm},\cite{Crane}-\cite{Chen} and references therein ), they still offer challenges even related
to the quantitative description of their topological structure.
At the dawn of the science of complex networks \cite{alb}, it is
therefore rather important to have a complete understanding of all
the tree structures and their properties.

A tree is defined as a set of points (vertices, nodes)
connected with line segments (branches, or edges) such that 
there are no cycles or loops (a connected graph without cycles).
For the simplest (unlabeled) rooted plane binary tree, each vertex 
has exactly three connecting branches, except for one vertex 
which is distinguished from all the others by having
only two connecting branches coined as the root
(R) of the tree, and a certain number of vertices
with a single connecting branch called the `leaves'. 
The height of the tree is defined by the maximum number of 
levels starting  from the root (which has height 0), and it can be 
calculated as the maximum number of branches one 
has to pass to reach the root from its
vertices (since the leaves have only one branch, it means that
this longest excursion must start from one of the leaves).
The paths from the leaves to the root define a natural
direction on the tree (similarly to the river flow) which
is always towards the levels of lower height.
A tree of height $k$ we call {\em complete}, if it has
$2^k$ leaves each being a distance  $k$
from the root. 

Let us now mention three applications of the mathematics
of trees  which
are  directly connected to the so-called Horton-Strahler
index of the tree, which is the subject of interest of the present 
paper.

Originally, the Horton-Strahler index of a binary tree 
was introduced in the studies of natural river networks by
Horton \cite{H} and later refined by Strahler \cite{S}, as a  
way of indexing real river topologies, since river networks 
are topologically similar to  binary trees. 
By definition, a leaf has a rank of 0 (some authors associate
the value of 1),
and a vertex has a rank of $r=r(i,j)$ where $r(i,j)$ is the
index function with $i$ and $j$ being the ranks of the
two connecting vertices from the level above. When
\begin{equation}
r(i,j)=\max{(i,j)}+\delta_{i,j},   \label{HSI}
\end{equation}
the index is called the 
Horton-Strahler index (HS). The quantity of particular interest 
is the HS index of the root which thus categorizes the topological
complexity of the whole tree. Several other quantities
can be introduced in relation to the HS index.
A {\em segment} of order $k$
\cite{VV}, or a {\em stream} of order $k$ \cite{Moon}
is a {\em maximal} path of
branches connecting vertices of HS index $k$,
ending in a vertex with index
$k+1$. Let ${\cal S}_k(n,T)$ denote the number of segments of order
$k$ of a tree $T$ with $n$ leaves, and $< L_k(n,T)>$ is the
average physical length of a segment of order $k$ (the average $<.>$
is taken on the tree $T$). The {\em bifurcation ratios}
${\cal B}_k(n,T)$ are defined as
${\cal B}_k(n,T) = {\cal S}_k(n,T) /{\cal S}_{k+1}(n,T)$, and the length
ratios via ${\cal L}_k(n,T) = < L_{k+1}(n,T)> / < L_{k}(n,T)>$.
Horton and
Strahler have empirically observed that for river networks both the
${\cal S}_k(n)$  and $ < L_k(n)> $ tend to approximate a 
geometric series, 
${\cal B}_k(n) \approx {\cal B}$ with $3 \leq {\cal B} \leq 5$
and ${\cal L}_k(n) \approx {\cal L}$ with $1.5 \leq {\cal L}
\leq 3$. Such networks are called {\em topologically self-similar}
\cite{Halsey}. The notion of HS index is further refined by introducing
the {\em biorder} $(i,j)$ of a vertex, representing the HS indices
of its two children \cite{Viennot1},\cite{VV}, \cite{Halsey}, and
then studying the ramification matrix, with elements 
related to the number of vertices with a given biorder.

Another interesting application of the mathematics of binary trees
and the HS index, is in the description of the branched structure
of diffusion-limited aggregates see Ref. \cite{Halsey} and references 
therein. In this case the structures are grown on a substrate
(which can be a point or a plane) by letting small particles diffuse
towards the aggregate where they stick indefinitely at their point
of first contact with the cluster. This creates complex and involved
branched structures, whose topological complexity still remains
a challenging problem to describe.

Finally, the last application we would like to mention 
is known as the {\em word bracketing problem} 
\cite{Comtet} which has obvious implications in computer science. 
Let us consider
an alphabet of $n$ letters, ${\cal A}=\{Y_1,Y_2,...,Y_n\}$ and a word
$S \equiv x_1 x_2 x_3 .. x_{n-1} x_n$, $x_i \in {\cal A}$. 
A 2-bracketing of the word $S$ is a partition of its letters
(by keeping their order) in groups of two units enclosed
in brackets, where a unit can be a letter or a subpartition 
enclosed in brackets, such as $(x_1x_2)(x_3(x_4x_5))$, or
$(x_1(x_2(x_3(x_4x_5)))$, etc. The bracket $(u_1 u_2)$ between
two units may be associated with a multiplicative composition
law $(u_1 u_2) = u_1 \cdot u_2 \in {\cal A}$. For example
let the alphabet ${\cal A}$ be all the positive integers,
and the composition law be the regular multiplication of numbers.
Then a bracketing of the multiple product $S$ corresponds to 
one particular way of calculating $S$. 
A one-to-one correspondence can be made immediately to trees:
let the letters ${x_1,x_2,...,x_n}$ of the word $S$ be associated
with the leaves of a binary tree. To a particular bracketing
of $S$ it corresponds a particular tree constructed by
associating a lower level vertex to a bracketing $(u_1 u_2)$
(one may think of the brackets as representing the
branches of the tree).
The main question is
 how many ways are there to calculate such a product.
If one assumes that the multiplication law is neither associative
nor commutative, then the problem is refered to as the Catalan
problem, see Ref. \cite{Comtet} for a number of solutions.
The number of such bracketings is given by the Catalan numbers,
$a_n = \frac{1}{n}{2n-2 \choose n-1}$. The corresponding set of 
trees (see Fig.1 for $n=4$) is in fact the set of rooted, unlabeled
binary plane trees according to this bijection.

\begin{figure}[htbp]
\protect\hspace*{2.0cm}
\epsfxsize = 5.0 in
\epsfbox{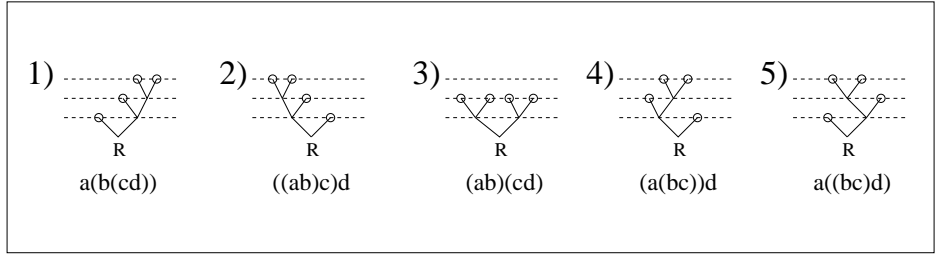}
\protect\vspace*{0.5cm}
\caption{The set of rooted, unlabeled
binary plane  trees corresponding to 
all the possible non-commutative, non-associative
bracketings of the four letter word $abcd$, $n=4$.}
\label{Fig1}
\end{figure} 

For later reference, we mention that the 
generating function $A(\xi)=
\sum_{n=0}^{\infty}\xi^n a_n$ 
of the Catalan 
numbers obey the equation $A^2-A+\xi =0$, with $A(0)=0$,
so 
\begin{equation}
A(\xi) = \frac{1}{2}(1-\sqrt{1-4\xi})\;. \label{A}
\end{equation}
The power series $A(\xi)$ converges within a disk 
of radius $a_c=1/4$.

The problem of enumerating trees becomes more difficult
if the composition law is commutative, which was
first studied by Wedderburn and Etherington (WE)
\cite{Wedderburn}, \cite{Etherington}, 
\cite{HP}. In the tree language,
this means that two trees are considered identical if 
after a number of successive reflections with respect to
the vertical axes passing through the vertices they can be
transformed into each other and in this case 
they are said to be homeomorphic \cite{Otter}. 
For the
example shown in  Fig. 1, there are only 
two such trees, since trees 1),
2), 4) and 5) can be transformed into each other. The 
trees that cannot be transformed into each other are
called non-homeomorphic. The set of non-homeomorphic
trees is called the set of {\em ambilateral} trees,
\cite{Smart}, \cite{Moon}.
Let the number of such
trees with $n$ leaves be denoted by $w_n$. The generating function
(GF) defined as $W(\xi)=\sum_{n=0}^{\infty} \xi^n w_n$
obeys
the nonlinear functional equation: 
\begin{equation}
W(\xi) = \xi+\frac{1}{2}W(\xi)^2+\frac{1}{2}W(\xi^2) \label{nlf}
\end{equation}
which has extensively been studied by Wedderburn 
\cite{Wedderburn}.  Otter \cite{Otter} 
studying a more
general counting problem where the vertices can have at most 
$m$ branches comes to the conclusion that for the 
ambilateral trees, if $n$ is
large we have: 
\begin{equation}
w_n \sim c \; n^{-3/2} \gamma^{n} \label{wen}
\end{equation}
where
$\gamma=2.4832535...$. The
method developed by Otter  gives an iterative approach
to $c$ and $\gamma$. For example $\gamma$ is 
\begin{equation}
\gamma = \lim_{n \to \infty} s_n^{2^{-n}} \label{iti1}
\end{equation}
where $s_0=2$, $s_n=2+s_{n-1}^2$ so that 
for $n=4$ one already obtains an extremely close
value of $\gamma \simeq (2090918)^{2^{-4}}$. Later, Bender
developed a more general approach \cite{Bender} 
deriving the same results. The coefficient $c$ in 
(\ref{wen}) can also be computed: $c = 0.31877662...$.

 The more practical application of the bracketing problem
within computer science is the computation of arithmetic
expressions by a computer. A general arithmetic expression
involving only binary operators can simply be mapped onto
a binary tree, called the syntax tree, which has as leaves
the operands and the inner vertices the operators. A computer
traverses this tree from the leaves towards the root and it
uses registers to store the intermediate results. In general
there are many ways of traversing such a tree, and the program
that uses the {\em minimal} number of registers is the most efficient,
or optimal one. Ershov has shown \cite{Ershov} that the optimal code
will use exactly as many registers to store the intermediate
results as the HS index of the associated syntax tree.

 In the present paper we investigate how the HS index is distributed
on both the rooted, unlabeled, plane binary set of trees, and on the
ambilateral set of binary  trees. We first answer this question on the
rooted, unlabeled, plane binary set, since it is simpler, but
it will also  provide us with a technique that can be extended
to tackle the problem for the ambilateral set.
For this set, the question was first answered
by Flajolet, Raoult and Vuillemin,  \cite{Flajolet} with a
method somewhat similar to the one presented here.
 The enumeration problem of the HS index on the
ambilateral set is, however, inherently more difficult since it involves
functional equations with nonlinear dependence in the argument
similar to Eq. (\ref{nlf}), and therefore an explicit solution in a
closed form becomes
impossible to attain. The derivation of an approximant formula
for the number of ambilateral trees sharing the same HS index at the
root is the main result of this paper.

The paper is organized as follows:
first we present our derivation of the enumeration
problem for the HS index
on the unlabeled set in Section II,
and then use this method of derivation
from this case to develop a technique that can be used  to attack 
the enumeration problem on the ambilateral set
in the asymptotic limit, presented in Section III. Section IV is
devoted to conclusions and outlook.

\section{Distribution of the HS index on the unlabeled set}

Let us observe that the root $R$ of the tree has always two subtrees
 attached to it  via the two branches, with $k$ and $n-k$ leaves,
respectively, $k=1,2,..,n$. Let $N^{(r)}_n$ denote the number
of unlabeled trees with $n$ leaves that share the same HS
index $r$ at the
root. A recursion is found for this number in the light of the
observation above:
\begin{eqnarray}
N^{(r)}_n = \sum_{k=1}^{n-1} \left\{ 
N^{(r-1)}_k  N^{(r-1)}_{n-k} + N^{(r)}_k 
\sum_{j=0}^{r-1} 
N^{(j)}_{n-k} + \right. 
\left. N^{(r)}_{n-k} \sum_{j=0}^{r-1}
N^{(j)}_k \right\} \label{Crec}
\end{eqnarray}
with the conventions $N^{(r)}_0 \equiv 0$, 
$N^{(0)}_n \equiv \delta_{n,1}$, $N^{(r)}_1 =\delta_{r,0}$.
If the generating function for the variable $n$  is defined
as $D_r(\xi) = \sum_{n=0}^{\infty} \xi^n N^{(r)}_n$, then
it obeys:
\begin{equation}
D_r = D_{r-1}^2 + 2 D_r \sum_{j=0}^{r-1} D_j\;,\;\;
r \geq 1,\;\;\;D_0=\xi \label{recC}
\end{equation}
Next we give an exact solution to (\ref{recC}). Let us introduce
the sum
$ %\begin{equation}
B_r \equiv \sum_{j=0}^{r-1} D_j,\;\;\; r \geq 1,
\;\;\;B_0=0,\;\;\;B_1=\xi\;.
$ %\end{equation}
Then $D_r=B_{r+1}-B_r$, and after rearranging the terms, Eq. 
(\ref{recC}) becomes $G_r=G_{r-1}$, where $G_r=B_r^2+
B_{r+1}(1-2B_r)$. This means that $G_r=G_0=\xi$, i.e.,:
\begin{equation}
B_r^2+B_{r+1}(1-2B_r) = \xi\;,\;\;\;r\geq 0 \label{B}
\end{equation}
Note that the left hand side of (\ref{B}) remains invariant
to  $B_r \mapsto 1-B_r$ which is another solution of (\ref{B}).
However, since in case of the HS index $B_0=0$, this latter
solution has to be dropped.
If we make $2 B_r \equiv 1-C_r$, (\ref{B}) simplifies to 
$C_r^2-2C_rC_{r+1}=4\xi-1$. Which, after dividing on both sides
by $4\xi-1$, and introducing $Z_r \equiv C_r/\sqrt{4\xi-1}$,
becomes:
\begin{equation}
Z_{r+1}=\frac{Z_r^2 -1}{2 Z_r} \label{Z}
\end{equation}
Let us now write $Z_r=\cot(v_r)$, such that $v_0 = 
\arctan{(\sqrt{4\xi-1})}$. 
Then (\ref{Z}) becomes
$\cot(v_{r+1})= \cot(2 v_r)$
which leads to $v_{r+1}=2v_r+\pi m$, 
$m \in \mathbb{Z}$, and which in turn is solved easily.
Thus, $Z_r = \cot(2^r v_0)$, so one finally obtains:
\begin{equation}
D_r(\xi) = \frac{\sqrt{4\xi-1}}{2 \sin\left(
2^{r+1} \mbox{arctg}\sqrt{4\xi-1}\right)}\;. \label{Ux}
\end{equation}
Eq. (\ref{Ux}) is the exact solution to (\ref{recC})
in the complex $\xi$ plane. 
On the real axis, within the radius of convergence
$a_c$ the above expression takes the form:
$ %\begin{equation}
D_r(\xi) = \sqrt{1-4\xi} \big/ \left[
2 \mbox{sh}\left(
2^{r+1} \mbox{arcth}\sqrt{1-4\xi}\right) \right]$,
$\xi < a_c=1/4 $.  %\end{equation}
Since within the convergence disk 
one must have $\sum_{r=0}^{\infty} D_r(\xi) = A(\xi)$,
we just obtained the identity (using (\ref{A})):
\begin{eqnarray}
1+\sum_{r=1}^{\infty} \frac{1}{\mbox{sh}(2^r x)} = 
\mbox{cth}{(x)},\;\;\;x > 0 \label{ident}
\end{eqnarray}
This identity can be checked to hold via more 
direct methods \cite{iza}.
The singularities of $D_r(\xi)$ lie on the positive
real axis at:
\begin{equation}
\xi_k^{(r)} = \frac{1}{4\cos^2 (k \pi/2^{r+1})},\;\;
k=1,...,2^r-1 \label{sing1}
\end{equation}
with an additional  singularity at infinity (corresponding to 
$k=2^r$). We certainly have $\xi_k^{(r)} > a_c$.
On the other hand if one simply iterates (\ref{recC})
we obtain:
\begin{equation}
D_r(\xi)=\frac{\xi^{2^r}}{2^r P_r(\xi)} \label{pol}
\end{equation}
where $P_r(\xi)$
is a polynomial in $\xi$ of order $2^r-1$: 
$P_1(\xi)=2^{-1}- \xi$,
$P_2(\xi) = P_1(\xi)(2^{-1}-2\xi+\xi^2)$,
$P_3(\xi) = P_2(\xi)(2^{-1}-4\xi+10\xi^2-8\xi^3+\xi^4)$,...
etc. 
One can find an explicit form for this polynomial from the
general solution (\ref{Ux}) if one invokes the identity
\cite{RG}:
$
\sin\left(n x\right) = n \sin x \cos x
\prod_{k=1}^{(n-2)/2} \left( 
1-\sin^2 x / \sin^2 (k \pi / n)
\right)
$,
so that (\ref{pol}) is recovered with:
$
P_r(\xi) = \prod_{k=1}^{2^r-1}
\mbox{ctg}^2(k\pi/2^{r+1})(\xi_k^{(r)}-\xi)
$.
It is easy to show, however, that $\prod_{k=1}^{2^r-1}
\mbox{ctg}^2(k\pi/2^{r+1}) = 1$, so the polynomial
simplifies to:
\begin{equation}
P_r(\xi) = \prod_{k=1}^{2^r-1}
(\xi_k^{(r)}-\xi) \label{prod}
\end{equation}
expression valid on the whole complex $\xi$ plane.
Based on the explicit 
solution we obtained, one can give an exact form to the
distribution of the HS index on the unlabeled set of trees,
by inverting the generating function via:
\begin{equation}
N^{(r)}_n = \frac{1}{2\pi i} \oint \frac{d\xi}{\xi^{n+1}}
D_r(\xi) = \frac{1}{2\pi i} \oint \frac{d\xi}{\xi^{n+1}}
\frac{\xi^{2^r}}{2^r P_r(\xi)}  \label{Cauchy}
\end{equation}
One can write:
\begin{equation}
\frac{1}{2^r P_r(\xi)} = \sum_{j=1}^{2^r-1}
\frac{A_j^{(r)}}{\xi_j^{(r)}-\xi} \label{apart}
\end{equation}
where
$
A_j^{(r)}=2^{-r}\prod\limits_{{k=1 \atop k \neq j}}^{2^r-1}
\left(\xi_k^{(r)}-\xi_j^{(r)}\right)^{-1}
,\;\;j=1,...,2^r-1
$.
By Cauchy's theorem the integrals in (\ref{Cauchy}) are 
readily performed, and one obtains:
\begin{eqnarray}
N^{(r)}_n = 
\left\{
\begin{array}{ll}
 \sum\limits_{j=1}^{2^r-1} 
 A_j^{(r)}\left[\xi_j^{(r)}\right]^{-(n-2^r+1)},\;
& n\geq 2^r \\
0, & \!\!\!\!\!0 \leq n \leq 2^r-1
\end{array} \right.\;\;\;\;\label{final0}
\end{eqnarray}
From (\ref{apart}) it follows that 
$
\sum_{j=1}^{2^r-1}A_j^{(r)}/\xi_j^{(r)} =
2^{-r}/ P_r(0) = 1
$.
To obtain the last equality we used the form (\ref{prod}) and
(\ref{sing1}). Thus:
$ %\begin{equation}
N^{(r)}_{2^r} = 1,\;\;\;r=1,2,...
$ %\end{equation}
The numbers $A_j^{(r)}$ can be calculated as follows: observe
that 
\begin{equation}
A_j^{(r)}=\lim_{\xi\to \xi_j^{(r)}} \frac{\xi_j-\xi}
{2^r P_r(\xi)} = - \frac{1}{2^r P_r'(\xi_j^{(r)})}, \label{aaj}
\end{equation}
where we used the L'H\^{o}pital rule in the last equality.
On the other hand from (\ref{pol}) and (\ref{Ux}) it follows:
$ %\begin{equation}
2^r P_r(\xi) = 2 \xi^{2^r}
\sin\left(
2^{r+1} \mbox{arctg}\sqrt{4\xi-1}\right) / \sqrt{4\xi-1}
$. %\end{equation}
Taking the derivative of this equation at the point $\xi_j^{(r)}$,
and inserting it in (\ref{aaj}) it yields:
\begin{equation}
A_j^{(r)}=(-1)^{j+1}\frac{4 \xi_j^{(r)}-1}{2^{r+1}
[\xi_j^{(r)}]^{2^r-1}}
\end{equation}
Thus, we obtain from (\ref{final0}) the more explicit form
\begin{equation}
N^{(r)}_n = \frac{1}{2^{r+1}}
\sum\limits_{j=1}^{2^r-1}
(-1)^{j+1}\frac{4 \xi_j^{(r)}-1}{[\xi_j^{(r)}]^n},
\;\;\;n\geq 2^r, \label{explicit}
\end{equation}
or using (\ref{sing1}):
\begin{equation}
N^{(r)}_n = \frac{4^{n}}{2^{r+1}}
\sum\limits_{j=1}^{2^r-1} 
(-1)^{j+1}\sin^2\left(\frac{j\pi}{2^{r+1}}\right)
\left[ \cos\left(\frac{j\pi}{2^{r+1}}
\right)\right]^{2n-2} \label{final1}
\end{equation}
an expression first derived by Flajolet et. al. \cite{Flajolet}.
Following this paper \cite{Flajolet}, our polynomials $P_r$ can be simply
connected to the Tchebycheff polynomial $U$ \cite{RG}, via the relation:
$P_r(\xi) = 2^{-r} \xi^{2^r - 1/2} U_{2^{r+1}-1}(1/(2\sqrt{\xi}))$.

If one employs the  Poisson resummation formula for functions defined
on a compact support (see Appendix B in Ref. \cite{toro}) on
(\ref{final1}), an
equivalent combinatorial expression can be derived in the form:
$N^{(r)}_{n+1} = \sum_{m=1}^{\infty} \nabla^2 \big[
{2n \choose n+k}\big](k)\big|_{k=1+(2m-1)2^r}$, where
$(\nabla f)(k) = f(k)-f(k-1)$ is the finite difference operator.
For a different method, see \cite{Flajolet}.

{\em Scaling limits.} Next we briefly present the results of
an asymptotic analysis on the $N^{(r)}_n$ numbers. Since  $N^{(r)}_n$
is an enumeration result, it typically contains several scaling
limits. In physical processes, during the growth of branched structures,
usually only one of these limits is selected, and in frequent cases
this limit has self similar properties (such as for DLA,
or for random generation of binary trees, \cite{bkm}).
By definition, the family of trees
that obey $\lim_{r \to \infty}(\ln{[n(r)]} ) / r =
const. \equiv \ln{{\cal B}}$ is
called topologically self similar \cite{Halsey}, where
${\cal B}$ is the {\em bifurcation number}. \\
1) $n \to \infty$ and $r$ fixed. In this case the first term in
(\ref{final1}) dominates the sum and the asymptotic behavior
is given by $N^{(r)}_n \sim 2^{-r-1} \mbox{tg}(\pi/2^{r+1})
e^{n \ln {[4 \cos^2 (\pi/2^{r+1})]}}$. The rate of the exponential
growth is a number between
 $\ln 2 $ and $ 2\ln 2$.
2) $n\to \infty$, $r \to \infty$, $\sqrt{n}/2^r \to \infty$. Here
the first term in (\ref{final1}) is still dominant (the rest being
exponentially small corrections) and yields:
$N^{(r)}_n \sim \pi^2 \; 2^{-3(r+1)}\;
e^{n \left( 2\ln 2 - \pi^2/4^{r+1} \right)}$. If $\sqrt{n}/2^r$
diverges with $r$ slower than exponential, we have topological self
similarity with ${\cal B} = 4$.\\
3) $n\to \infty$, $r \to \infty$, $\sqrt{n}/2^r \to d$,
with some $0 < d < \infty$. In this case the rest of the terms in
(\ref{final1}) (after the first has been factored out) are of the
type $j^2 e^{-(j-1)\pi^2 d^2}$ and the final expression is:
$N^{(r)}_n \sim A(d)\; 4^n n^{-3/2}\;4^n$. The topological
self similarity is obvious with ${\cal B} = 4$. The factor $A(d)$
is given by $A(d)= \pi^2 e^{-\pi^2 d^2} (1-e^{-\pi^2 d^2})/
(1+e^{-\pi^2 d^2})^3$.\\
4)$n\to \infty$, $r \to \infty$, $\sqrt{n}/2^r \to 0$, and
$n/2^r \to \infty$. In this case the analysis is performed easier
from the combinatorial expression of $N^{(r)}_n$ and yields:
$N^{(r)}_n \sim \pi^{-1/2}\;n^{-5/2}\;e^{n 2\ln 2 -4^r/n}$.

\newpage

\section{Distribution of the HS index on the ambilateral set}

Let us now analyze the same question on the set of
ambilateral trees, and denote the number of ambilateral trees with
$n$ leaves and HS index $r$ by $M^{(r)}_n$.
We certainly must have the
relation \begin{equation}
\sum_{r=0}^{\infty} M^{(r)}_n = w_n\;. \label{closure}
\end{equation}

\begin{figure}[htbp]
\protect\vspace*{-0.5cm}
\protect\hspace*{4.0cm}
\epsfxsize = 3.0 in
\epsfbox{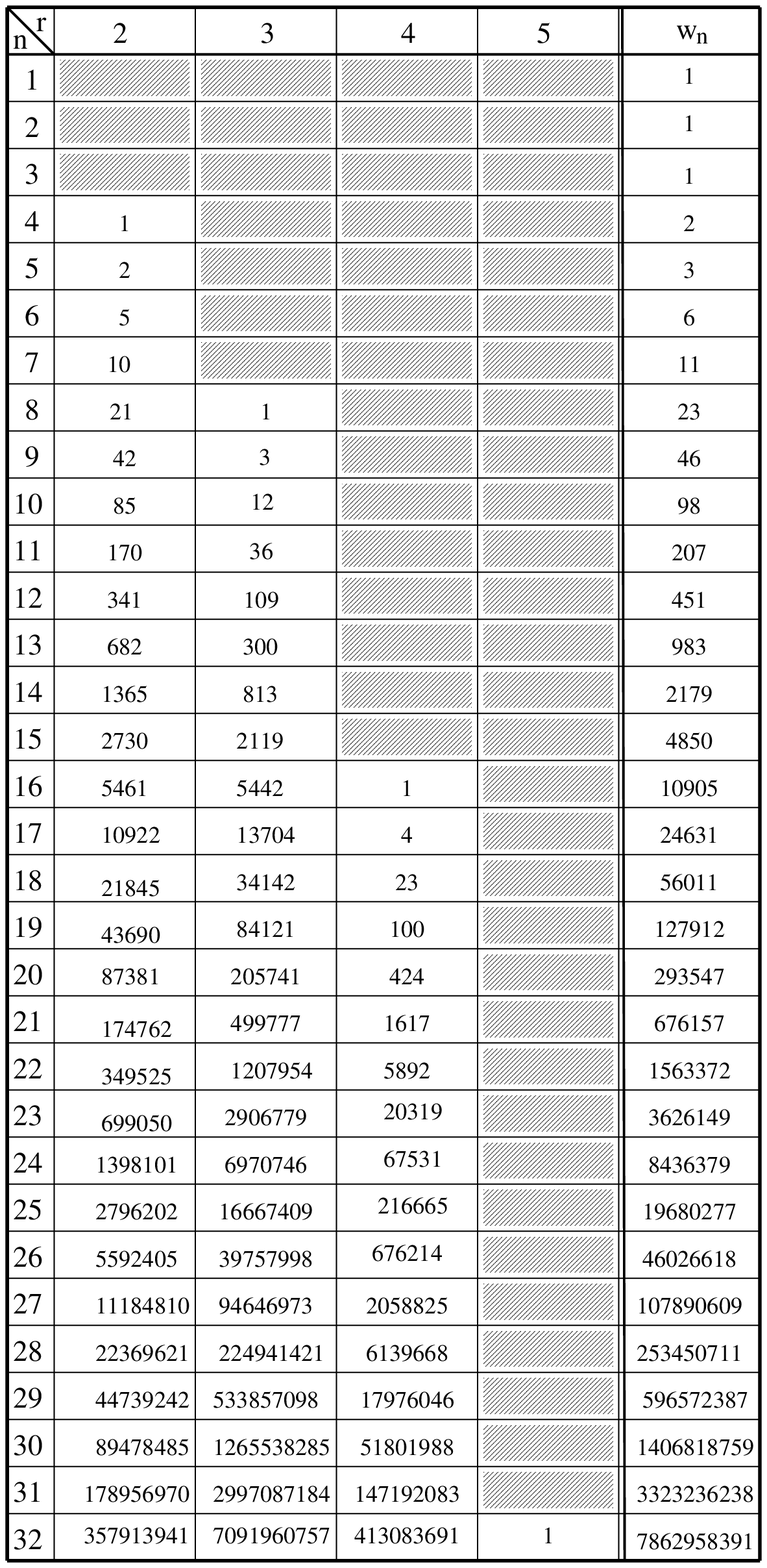}
\protect\vspace*{-0.0cm}
\caption{Particular values for the number of ambilateral trees with $n$ leaves and HS index $r$.
The shaded entries mean that there is no such tree.} \label{tbl1}
\end{figure}

 The table in Fig. \ref{tbl1} gives the distribution of 
the HS index for $n$ up to 32 and $r=2,3,4,5$. We can check easily that
$M^{(0)}_n = \delta_{n,1}$, and $M^{(1)}_n = 1-\delta_{n,1}$, 
so for simplicity  these are not represented in the table.

The numbers $M^{(r)}_n$ obey slightly more complicated 
recurrence  relations since now the counting has to be done on a more
restricted set. 
We must  distinguish between odd and even $n$ values. However,
the two cases can be combined into one, if the convention
$M^{(r)}_{\nu}=0$ for $\nu$ non-integer is adopted. The
corresponding recurrence relation becomes:
\begin{eqnarray}
M^{(r)}_n = \!\!\!\sum_{0\leq k < j \leq n}\!\!\!
\delta_{k+j,n} \Bigg[ 
M^{(r-1)}_k M^{(r-1)}_j + 
 \sum_{s=0}^{r-1} \left( 
M^{(r)}_k M^{(s)}_j
+M^{(s)}_k M^{(r)}_j\right)\Bigg]+
M^{(r)}_{n/2} \sum_{s=0}^{r-1} M^{(s)}_{n/2}
+\frac{1}{2} M^{(r-1)}_{n/2}\Big( 
1+M^{(r-1)}_{n/2}\Big)
\end{eqnarray}
The generating function 
$V_r(\xi) = \sum_{n=0}^{\infty} \xi^n M^{(r)}_n$
will thus obey:
\begin{equation}
V_r(\xi) = \frac{1}{2}\;
 \frac{\big[V_{r-1}(\xi)\big]^2 +
V_{r-1}(\xi^2)}
{1-\sum\limits_{s=0}^{r-1} V_s(\xi)},
\;\;r\geq 1, \label{rec2}
\end{equation}
and $V_0(\xi) = \xi$. 
As a check for the correctness of (\ref{rec2}), let us see if we
recover the identity $\sum_{r=0}^{\infty} V_{r}(\xi) = W(\xi)$
(which follows from (\ref{closure})). Eq. (\ref{rec2}) is equivalent
to $2V_r(\xi) - 2 \sum_{s=0}^{r-1} V_{s}(\xi)V_{r}(\xi) = 
\big[V_{r-1}(\xi)\big]^2 +
V_{r-1}(\xi^2) $. Introduce the temporary variable $G(\xi) = 
\sum_{r=0}^\infty V_r(\xi)$ and sum both sides of the equation
over $r$, $r=1,2,..,\infty$. One obtains $2(G(\xi)-\xi) -2 
\sum_{r=1}^{\infty}\sum_{s=0}^{r-1}V_{s}(\xi)V_{r}(\xi)  = 
\sum_{r=0}^\infty \big[V_r(\xi)\big]^2 + G(\xi^2)$. Using the
identity $2 \sum_{r=1}^{\infty}\sum_{s=0}^{r-1}V_{s} \;V_{r}
 = \left(\sum_{r=0}^{\infty} V_r\right)^2 - \sum_{r=0}^{\infty} V_r^2$,
one finds $G(\xi) = \xi + \frac{1}{2}\big[G(\xi)\big]^2 + 
\frac{1}{2} G(\xi^2)$ which is precisely Eq. (\ref{nlf}), showing that
$G(\xi) = W(\xi)$, i.e., the relation $\sum_{r=0}^{\infty} V_{r}(\xi) 
= W(\xi)$ holds, indeed.

\begin{figure}[htbp]
\vspace*{-0.3cm}
\protect\hspace*{3.5cm}
\epsfxsize = 4.0 in
\epsfbox{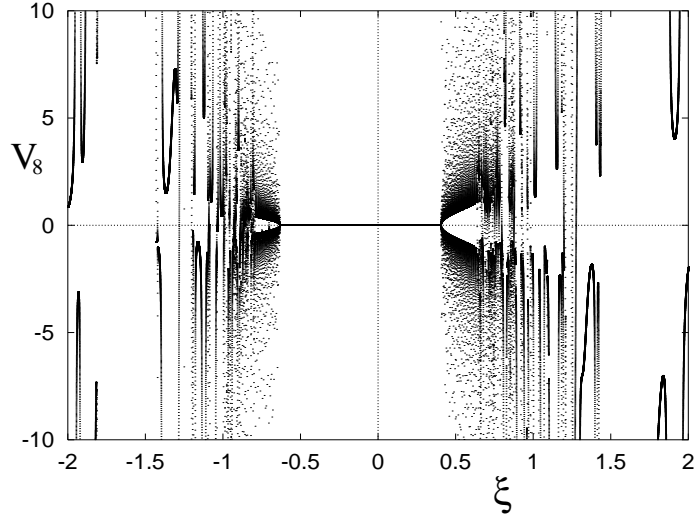}
\protect\vspace*{0.5cm}
\caption{The generating function $V_8(\xi)$ on the
real axis. The function was evaluated in more than 
$1.3\cdot 10^5$ points, and represented by dots.} \label{Fig4}
\end{figure}

In contrast to the previous case,
the functional recurrence (\ref{rec2}) cannot be treated
in an exact analytical fashion due to the
functional dependence on $\xi^2$. However, one can 
derive the asymptotic behaviour and make statements
that will lead to rather close approximations 
of the $M^{(r)}_n$ numbers. It is instructive to look at a 
few particular values, first:
\begin{eqnarray}
&& V_1(\xi) = \frac{\xi^2}{1-\xi},\nonumber \\
&&V_2(\xi) = \frac{\xi^4}{(1-2\xi)(1-\xi^2)}, \\
&& V_3(\xi) =\xi^8 \frac{1-2\xi+\xi^2+2\xi^3-
3\xi^4}{(1-2\xi)(1-2\xi^2)(1-\xi^4)
(1-3\xi+4\xi^3-\xi^4)}\;.\nonumber
\end{eqnarray}
Inverting $V_2(\xi)$, one obtains:
$M^{(2)}_n=
\big[ 2^{n-1}-3+(-1)^{n-4}\big]/6$, $n \geq 4$,
which can be checked to hold, see the table in Fig.
\ref{tbl1}. The result from the inversion
of $V_3(\xi)$ is already so complicated that it is 
not worth presenting.
As the index $r$ increases, the polynomial expressions become
more and more involved. Figure \ref{Fig4} shows the function
$V_8(\xi)$ in the interval $[-2.0,2.0]$.

For every $r$, the power series for $V_r(\xi)$ has non-negative
coefficients, $M^{(r)}_n \geq 0$. Based on a classic theorem 
of complex analysis, this means that 
on the circle of convergence, 
of radius $\alpha_r > 0$, there will be a singularity of $V_r(\xi)$
at $\xi = \alpha_r$. Next we show, that we have the ordering
$0 < \alpha_{r+1} < \alpha_r < 1$ for $r \geq 2$, and the
limit $\lim_{r \to \infty} \alpha_r$ exists and it is equal to 
$\alpha \equiv 1/\gamma = 0.4026975036...$.
We shall use mathematical induction
to prove the ordering. From the particular examples above it
follows that $\alpha_2 = 0.5$, $\alpha_3 = 0.424507...$.
Let us now assume that $\alpha_{j} < \alpha_{j-1} < 1$ for all $j \leq r$,
$j \geq 2$.
We will show that $\alpha_{r+1} < \alpha_r$. 
Note that the radius of convergence for $V_j(\xi^2)$ is $\sqrt{\alpha_j}
> \alpha_j$, if $\alpha_j$ is less than unity.
By {\em reductio ad absurdum},
let us assume first, that $\alpha_{r+1} > \alpha_r$. This means, that
$V_{r+1}(\xi)$ is analytic in $\alpha_r$.  From (\ref{rec2}),
\begin{equation}
V_{r+1}(\xi) =
\frac{V^2_r(\xi) + V_r(\xi^2) }
{1-V_0(\xi)-...-V_r(\xi)}\;. \label{4rp1}
\end{equation}
By the argument above, $V_r(\xi^2)$ is analytic in $\alpha_r$ (its
radius of convergence is $\sqrt{\alpha_r} > \alpha_r$, since by assumption
$\alpha_r < \alpha_{r-1} < ... < \alpha_2 = 1/2 < 1$). In the
denominator of (\ref{4rp1}),
{\em all} functions $V_j$, $j=0,1,.. r-1$ are analytic in
$\alpha_r$, because by assumption  they all have radii of convergence 
strictly larger than $\alpha_r$.  However, $V_r$ is singular in 
$\alpha_r$, and
the singularities do not cancel in the numerator and 
denominator of (\ref{4rp1}), and thus
$V_{r+1}$ is singular in $\alpha_r$, a contradiction.
We are left to prove
that  $\alpha_{r+1} =  \alpha_r$ cannot hold. Let us denote $B_r = 
\sum_{s=0}^{r} V_s$.
Again, we assume, that 
$\alpha_{r+1} =  \alpha_r$ is true. It is easy to show, that for
any {\em finite} $r$, $|V_r(\alpha_r)| = \infty$. This means from
the recurrence relation that 
\begin{equation}
B_{r-1}(\alpha_r) = 1 \label{aneq}
\end{equation} 
(in the numerator
of (\ref{rec2}) we have only functions analytic at $\alpha_r$).
Since $V_{r+1}(\xi) = 
\left[V^2_r(\xi) + V_r(\xi^2) \right]/[1-B_r(\xi)]$, from
the assumption $\alpha_{r+1} =  \alpha_r$ it would follow that
the equation $B_r(x) = 1$ {\em cannot} have any solutions ($V_{r+1}$
is analytic within the circle of convergence)
in the interval $0 < x < \alpha_r$. (Note that in the interval 
$0 < x < \alpha_r$, the numerator $V^2_r(\xi) + V_r(\xi^2)$ 
cannot be zero, since the power series $V_r$ has only 
positive coefficients).
The equation $B_r(x) = 1$ is equivalent to $B_{r-1}(x) + 
V_r(x) = 1$. However,
from (\ref{rec2}) $1-B_{r-1}(x) = \left[V^2_r(x) + 
V_r(x^2)\right]/V_r(x)$,
Thus, the equation
\begin{equation}
V^2_r(x) = V^2_{r-1}(x) + V_{r-1}(x^2) \label{equ}
\end{equation}
should have no solution in $0 < x < \alpha_r$. If $x$ is arbitrarily
close to $\alpha_r$, then $V^2_r(x)$ is arbitrarily large. However,
since $\alpha_{r-1} > \alpha_r$, $V^2_{r-1}(x)$, and $V^2_{r-1}(x^2)$ 
are both bounded from above. Thus, for $x$ sufficiently  
close to $\alpha_r$, we have $V^2_r(x) >  V^2_{r-1}(x) + V_{r-1}(x^2)$.
On the other hand, the HS index of a tree $T$ equals to the
height of the largest, complete, balanced tree embedded in $T$. This
means, that $M^{(r)}_n = 0$ for $n = 0,1,2,..,2^r-1$. Also, 
$M^{(r)}_{2^r} = 1$. In other words, one must have $V_r(x)
=x^{2^{r}}\left(1+{\cal O}(x) \right)$. 

\begin{figure}[htbp]
\protect\hspace*{3.5cm}
\epsfxsize = 4.0 in
\epsfbox{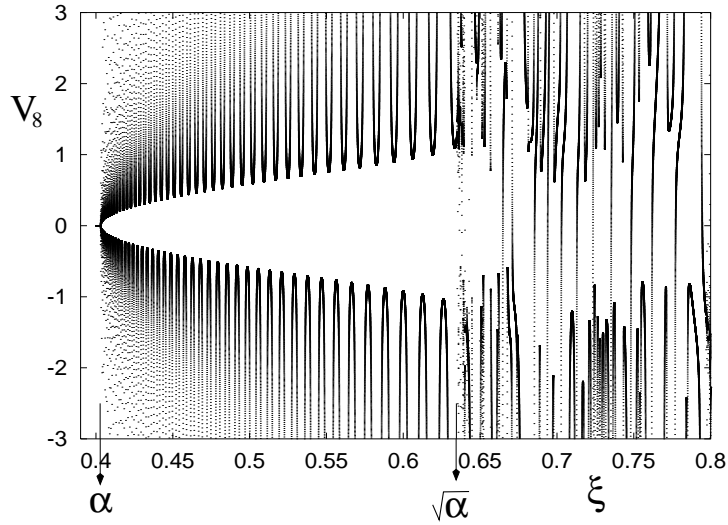}
\protect\vspace*{0.5cm}
\caption{A magnified region of Figure \protect\ref{Fig5}
The arrows indicate the positions $\alpha = 0.40269...$ and 
$\sqrt{\alpha} = 0.63458...$ on the
real axis.} \label{Fig5}
\end{figure}

This means that
$V^2_{r-1}(x) = x^{2^r} \left(1+{\cal O}(x) \right)$,
$V_{r-1}(x^2) = x^{2^r} \left(1+{\cal O}(x^2) \right)$, and
$V^2_r(x) = x^{2^r} x^{2^r} \left(1+{\cal O}(x) \right)$.
Since $V^2_{r-1}(x) + V_{r-1}(x^2) = 2 x^{2^r} \left(1+{\cal O}(x) 
\right)$, there will always be an $x > 0$, ($x < 1$), sufficiently
close to zero, such that  $V^2_r(x) <  V^2_{r-1}(x) +
V_{r-1}(x^2)$. Therefore, there must exist an $0 < x < \alpha_r$,
for which
(\ref{equ}) holds, which is a contradiction. Thus, we have
proven that $0 < \alpha_{r+1} < \alpha_r < 1$, for all $r \geq 2$.
As a matter of fact we have also shown, that the radii of convergence
satify:
\begin{equation}
V^2_r(\alpha_{r+1}) = 
V^2_{r-1}(\alpha_{r+1}) + V_{r-1}(\alpha_{r+1}^2),\;\;\;r\geq 1.
\label{equ1} \end{equation}
Since the series $\alpha_r$ is monotonically decreasing, and bounded
from below, the limit $\alpha = \lim_{r \to \infty} \alpha_r$ exists. 

We have shown that $\sum_{r=0}^{\infty} V_r{(\xi)} = W(\xi)$. 
Since the radius of convergence for the left hand side is the minimum
of all the radii of the terms in the summation, i.e., $\alpha$,
it must equal to the radius of convergence
for $W(\xi)$, which, as shown by Otter and Bender is $1/\gamma$,
$\lim_{r \to \infty} \alpha_r = \alpha = 1/\gamma = 0.4026975036...$. 
Taking the limit $r \to \infty$ in (\ref{aneq}), we get 
\begin{equation}
W(\alpha) = 1   \label{wg}
\end{equation}
(since by definition $B_r = \sum_{s=0}^{\infty} V_s$, so 
$\lim_{r \to \infty} B_r = W(\xi)$). Or, using (\ref{nlf}):
\begin{equation}
W(\alpha^2) = 1-2\alpha  \label{iter}
\end{equation}
an identity also shown by Bender. Eqs. (\ref{iter}) and (\ref{nlf})
can simply be combined to give the iterative computation of $\alpha$
in the form already mentioned in the Introduction, as follows: if
we make the temporary notation 
\begin{equation}
U(\xi) \equiv 
[1-W(\xi)]/\sqrt{\xi}\;, \label{UU}
\end{equation}
Eq. (\ref{nlf}) takes the form
\begin{equation}
U(\xi^2) = 2+U^2(\xi)\;, \label{si1}
\end{equation} 
and Eq. (\ref{iter}) simplifies to 
\begin{equation}
U(\alpha^2) = 2\;. \label{si2}
\end{equation} 
Let $S(x) = 2+x^2$. Then, from (\ref{si1}) $U(\xi^2) = S(U(\xi))$,
or $U(\xi) = S(U(\xi^{1/2}))=S(S(U(\xi^{1/4}))) = 
S(...S(U(\xi^{2^{-n}}))...)$, where there are a total of $n$ compositions
for the $S$ function, $n$ arbitrary. 
Let us now choose $\xi = \alpha^{2^{n+1}}$. This means, 
$U(\alpha^{2^{n+1}}) =
S(...S(U(\alpha^2))...) = S(...(S(2))...)$, by virtue of (\ref{si2}).
From (\ref{UU}), $U(\alpha^{2^{n+1}}) =
[1-W(\alpha^{2^{n+1}})]/\alpha^{2^{n}}$. We have shown previously, that
$\alpha < 1$ (it is the limit  of the monotonically
decreasing series $\alpha_r <1$), therefore we have: 
\begin{equation}
\alpha = \lim_{n \to \infty} \left(
\frac{1-W(\alpha^{2^{n+1}})}{s_n}
\right)^{2^{-n}} = \lim_{n \to \infty} s_n^{-2^{-n}} \label{afa}
\end{equation}
since $W(\alpha^{2^{n+1}}) \to W(0) = 0$, and where $s_n = S(s_{n-1})$, $s_0 =
2$, just as in the Introduction. The convergence is double-exponential,
very fast.

\begin{figure}[htbp]
\protect\hspace*{3.5cm}
\epsfxsize = 4.0 in
\epsfbox{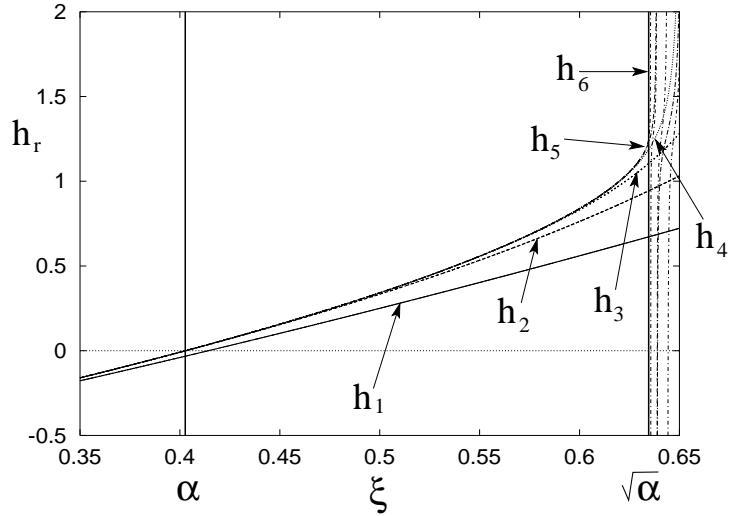}
\protect\vspace*{0.5cm}
\caption{The functions $h_r(\xi)$ are analytic on $I$. This figure
shows $h_r(\xi)$ for $r=1,2,3,4,5,6$. 
The convergence on $I$ to $h(\xi)$ is double-exponentially fast.
The thick vertical lines delimit the edges of the interval $I$.
Close to $\alpha$, the $h_r$ functions cannot be distinguished on $I$ for $r \geq 3$.
To the right from $I$ the $h_r$ functions develop singularities.
The point $\sqrt{\alpha}$ is a left accumulation point for the
series of the leftmost singularities of $h_r(\xi)$ as $r \to \infty$.}
\label{Fig6} \end{figure}

As in Section II, the asymptotic behavior of the
$M^{(r)}_n$ numbers for relatively large $n$ and $r$ is governed
by the innermost singularity of $V_r(\xi)$ on the real axis.
The graph of $V_8$ shown in Figure \ref{Fig4} suggests, that the
generating function is in fact well behaved in a certain interval
to the right of the radius of convergence, $\alpha_8$, 
see also Figure \ref{Fig5}. The existence of this interval comes from
the fact that the singularities of the term with nonlinear argument 
$V_{r-1}(\xi^2)$ in
the numerator of (\ref{rec2}) kick in only beyond the circle of convergence
of $V_{r-1}(\xi^2)$, which is $\sqrt{\alpha_{r-1}} > \alpha_{r-1}$.
Thus, in the interval $\alpha_r < x < \sqrt{\alpha_{r-1}}$ 
the term with the nonlinear argument is analytic, which ultimately is
responsible for this nice behaviour. Because, $\alpha <
\alpha_{r-1}$, for convenience we shall define the interval of this nice
behaviour to be $I  =  [\alpha, \sqrt{\alpha})$. In order to exploit this
observation, we shall first rewrite the recurrence relation (\ref{rec2}).
Let us denote $H_r(\xi) = 1-\sum_{s=0}^{r-1} V_s(\xi)$. With this
notation, (\ref{rec2}) takes the form $G_r(\xi) = G_{r-1}(\xi)$, 
$r\geq 1$, where $G_r(\xi) = H^2_r(\xi)-2H_r(\xi)H_{r+1}(\xi)+
H_r(\xi^2)$. This leads to the new recurrence:
\begin{equation}
H^2_r(\xi)-2H_r(\xi)H_{r+1}(\xi)+
H_r(\xi^2) = 2\xi\;, \label{rec3}
\end{equation}
$H_0(\xi) = 1$. This would be exactly solvable if it were not for the
dependence on the nonlinear argument $\xi^2$. Note the resemblance to
(\ref{B}). Let $h_r(\xi) = 2\xi - H_r(\xi^2)$, which is an
analytic function
in $I$. We also have
$ %\begin{equation}
\Delta h_r(\xi) = h_r(\xi) - h_{r-1}(\xi) = V_r(\xi^2) =
\xi^{2^r} \left(1+{\cal O}(\xi^2) \right)
$, %\end{equation}
the latter equality being shown previously. This shows, that in the
interval $I$, the $r$-dependence {\em weakens extremely fast}, 
double-exponentially with 
increasing $r$.
As a matter of fact, an upper estimate is 
\begin{equation}
\Delta h_r(\xi) \leq \alpha^{2^{r-1}}\;.  \label{diffu}
\end{equation}
In particular, 
$\Delta h_3(\xi) \leq 0.0263$, 
$\Delta h_4(\xi) \leq 0.0006916$, 
$\Delta h_5(\xi) \leq 4.79\cdot 10^{-7}$,
$\Delta h_6(\xi) \leq 2.28\cdot 10^{-13}$, 
$\Delta h_7(\xi) \leq 5.22\cdot 10^{-26}$, etc. Therefore, from the
point of view of the asymptotic behavior, the $h_r$ functions
can be replaced by their asymptotic expression (as $r \to \infty$):
\begin{equation}
h(\xi) = W(\xi^2) + 2\xi -1\;. \label{smooth}
\end{equation}
Figure \ref{Fig6} shows the functions $h_r$ on the interval $I$ for
$r=1,2,3,4,5,6$.

Thus, instead of Eq. (\ref{rec3}) we will consider:
\begin{equation}
\overline{H}^2_r(\xi)-2\overline{H}_r(\xi)\overline{H}_{r+1}(\xi) 
= h(\xi)\;. \label{app} 
\end{equation}
The recurrence (\ref{app}) in turn is easily solved in the 
way shown in Section I. The result is:
\begin{equation}
\overline{H}_r(\xi) = \sqrt{h(\xi)}\;\mbox{ctg}\!\left(
2^{r-r_0} \mbox{arctg}\left(\frac{\sqrt{h(\xi)}}
{\overline{H}_{r_0}(\xi)}\right)\right) \label{hrs}
\end{equation}
where $r_0$ for the moment is an arbitrary (positive integer) index.
Recurrence (\ref{app}) will become a good approximation to the
recurrence (\ref{rec3}) from an index $r_0$ on. The larger $r_0$
is the more accurate the approximation. Recurrence (\ref{app}) is
applied then with initial condition $\overline{H}_{r_0}(\xi) = 
H_{r_0}(\xi)$, which for modest $r_0$ values can be obtained
by iterating (\ref{rec3}) $r_0$ times. 

What is the error we make when one replaces $h_{r_0}(\xi)$ with 
$h(\xi)$ on $I$? Summing the differences (\ref{diffu}) from
$r_0 + 1$ to infinity, one obtains the estimate:
$h(\xi) - h_{r_0}(\xi) \leq \alpha^{2^{r_0}}\sum_{m=0}^{\infty}
\alpha^{2^{r_0} (2^{m}-1)} <
\frac{\alpha^{2^{r_0}}}{1-\alpha^{2^{r_0}}}$.
Thus, for example, $h(\xi) - h_{5}(\xi)$ is smaller than $10^{-7}$,
$h(\xi) - h_{6}(\xi)$ is smaller than $10^{-13}$, etc.

Therefore, we can finally write on $I$:
\begin{equation}
V_r(\xi) \simeq \frac{\sqrt{W(\xi^2)+2\xi-1}}{\sin{\left( 
2^{r+1-r_0} \mbox{arctg}\left(\sqrt{W(\xi^2)+2\xi-1}/
H_{r_0}(\xi)\right)
\right)}}\;,
\;\;\;r > r_0 \;,\;\;\;\xi \in I\;. \label{apo}
\end{equation}

\begin{figure}[htbp]
\protect\hspace*{3.5cm}
\epsfxsize = 4.0 in
\epsfbox{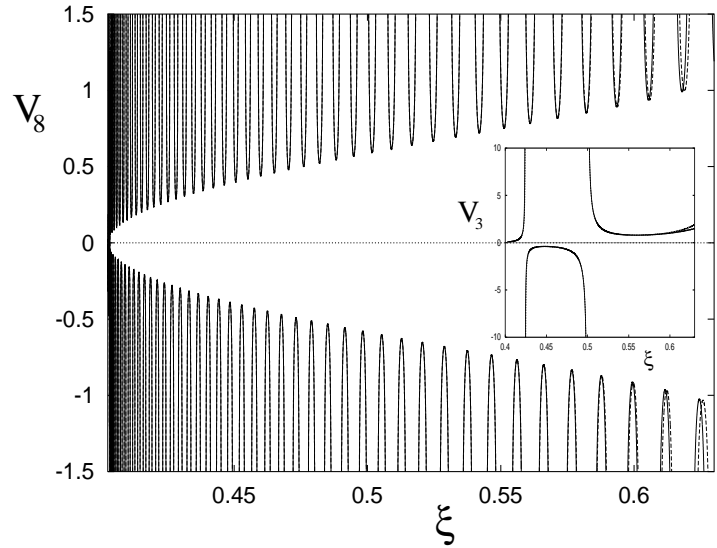}
\protect\vspace*{0.5cm}
\caption{The true $V_r(\xi)$ function (dashed line) from iterating
(\ref{rec2}), and the approximation in Eq. (\ref{apo})
(solid line) for $r=8$
with $r_0 = 3$, and $r=3$ with $r_0 = 2$
(the inset). } \label{Figxy}
\end{figure}

In Fig. (\ref{Figxy}) we plot the rhs of (\ref{apo}) and the $V_r$
function from iterating (\ref{rec2}). Note that the approximation
is very good, and it becomes virtually indistinguishable from the
true function the closer $\xi$ is to $\alpha$. Larger $r_0$ values
will also give better approximations, since the approximation
is only applied from the $r_0$ index on. However, $r_0$ cannot be taken
too high for approximation purposes, since it assumes that the
exact expression of $H_{r_0}$ (or $V_{r_0}$) is known. 
This makes only the modest $r_0$ values (less than 5) useful.
On the other hand, expression (\ref{apo}) is very practical in analysing
the singularities of $V$ and give rather close approximant expressions
to these singularities. In particular, we see that within the interval
$I$, (\ref{apo}) preserves the property that if $\alpha_{r'}$ is
a singularity of $V_{r'}$ (or a zero of $H_{r'}$) then it is a singularity
of $V_{r}$ (or a zero of $H_{r}$), whenever $r > r'$. If one is interested
in the asymptotic behavior, then a more tractable expression can be 
derived for the rhs of (\ref{apo}): the function $h(\xi)$ is analytic on 
the interval $I$, and since already for modest $r$ values, the innermost
singularity of $V_r$ (denoted $\alpha_r$) is extremely close 
to $\alpha$, one can safely replace $h(\xi)$ in this neighborhood
by: $h(\xi) \simeq h'(\alpha) (\xi-\alpha)$.

\begin{figure}[htbp]
\protect\hspace*{3.5cm}
\epsfxsize = 4.0 in
\epsfbox{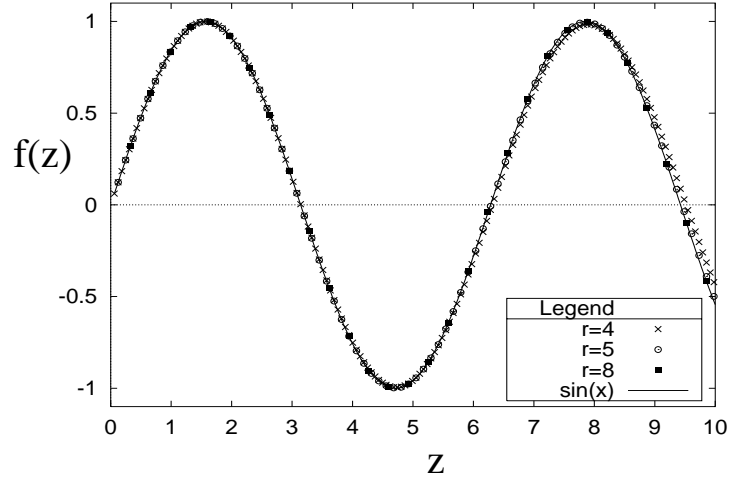}
\protect\vspace*{0.5cm}
\caption{The goodness of (\ref{apo1}). For $\mu$ and $\theta$
we used the values derived in the text.} \label{fig9}
\end{figure}

This leads to the approximant:
\begin{equation}
V_r(\xi) \simeq K_r(\xi) \equiv \frac{\mu \sqrt{\xi-\alpha}}{\sin{\left( 
2^{r+1} \mbox{arctg}\left(\theta \sqrt{\xi-\alpha}\right)
\right)}}\;,\;\;\;\xi \in I\label{apo1}
\end{equation} 
for sufficiently large $r$ (here ``large'' means $r\geq 4$) where
\begin{equation}
\mu  =  \sqrt{h'(\alpha)}\;,\;\;\;\;
\theta = \frac{\sqrt{h'(\alpha)}}{2^{r_0}H_{r_0}(\alpha)}\;.
\end{equation}
Next, we compute $h'(\alpha)$. One can use a very similar method to the
one employed to obtain (\ref{afa}), to give:
\begin{equation}
h'(\alpha) = \lim_{n \to \infty} \frac{s_n}{s_0 s_1 ... s_{n-1}}
= 3.1710556... \label{deriv}
\end{equation}
so, $\mu = 1.780745815...$.
If one computes $\theta$ for $r_0 = 3$, we have 
$H_3(\alpha) =(1-3 \alpha + 4 \alpha^3 -\alpha^4)/(1-2 \alpha
-\alpha^2 + 2 \alpha^3) = 0.164518..$, and thus $\theta = 1.3530022...$.
If we were to use $r_0 = 4$, then one would obtain $H_4(\alpha) = 
0.082262$, so $\theta = 1.3529529245$ and slightly improve 
the approximation on $\theta$.
No significant improvement will be obtained with larger $r_0$ values.
Figure \ref{fig9} shows the agreement of the form given in (\ref{apo1}).
For clarity, we defined the function $f(z)$ given by:
\begin{equation}
f(z) = \frac{\mu}{\theta} 
\frac{ \mbox{tg} \left( \frac{z}{2^{r+1}} \right) } 
{V_r \left( \alpha + \theta^{-2}
\mbox{tg}^2\left( \frac{z}{2^{r+1}} \right) \right)}
\end{equation}
Here we use the true $V_r$ function using numerical iteration of
(\ref{rec2}), and evaluate it in the points $\xi = \alpha + \theta^{-2}
\mbox{tg}^2\left( \frac{z}{2^{r+1}} \right) $. If the approximation
(\ref{apo1}) is good, then one should have $f(z) = sin(z)$. As seen
from Fig.~\ref{fig9} the approximation is already excellent for $r=4$
close to $\alpha$ (which corresponds to the $z=0$ point in this plot).
The interval $I$ in these transformed corrdinates corresponds
to $(0, 2^{r+1} \mbox{arctg}(\theta \sqrt{\sqrt{\alpha} - \alpha})) = 
(0, 0.577435486 \cdot 2^{r+1})$.
There are no fitting parameters, we used for $\mu$ and 
$\theta$ the values derived above.

In order to obtain the approximation to the {\em number}
$M^{(r)}_n$ of ambilateral trees with the same HS index at the root, 
we will have to invert (\ref{apo1}). The singularities of the rhs of
(\ref{apo1}) are given by: 
\begin{equation}
\xi^{(r)}_k = \alpha + \theta^{-2} \mbox{tg}^2\left( 
\frac{k\pi}{2^{r+1}}\right)\;,\;\;\;k=1,2,3,...,2^r-1 \label{sing2}
\end{equation}
(at the moment we do not care whether some of the singularities
will fall outside the interval $I$, we just simply want to invert
(\ref{apo1}), and then at the end keep only those terms from the
final expression that were generated by the singularities within
$I$).

 In a similar manner to the previous section, we first
bring $K_r$ to an inverted polynomial form:
\begin{equation}
K_r(\xi) = \frac{\mu \left[1+\theta^2(\xi-\alpha)\right]^{2^r}}
{2^{r+1} \theta^{2^{r+1}-1} Q_r(\xi)}
\end{equation}
where $Q_r$ is the polynomial:
$Q_r(\xi) = \prod_{k=1}^{2^r-1} \left( \xi^{(r)}_k - \xi \right)$.
The case from the previous Section II corresponds
to $\mu=1, \theta = 2$ and $\alpha = 1/4$.  Thus, if we denote
by $\overline{M}^{(r)}_n$ the numbers coming from the inversion
of $K_r(\xi)$, then:
\begin{equation}
\overline{M}^{(r)}_n = \frac{\mu}{ \theta^{2^{r+1}-1}}
\frac{1}{2\pi i} \oint \frac{d\xi}{\xi^{n+1}} 
\frac{\left[1+\theta^2(\xi-\alpha)\right]^{2^r}}
{2^{r+1} Q_r(\xi)}
\end{equation}
We have:
\begin{equation}
\frac{\mu}{2^{r+1} \theta^{2^{r+1}-1}Q_r(\xi)} = \sum_{j=1}^{2^r-1} 
\frac{A^{(r)}_j}{\xi^{(r)}_j - \xi},\;\;\;
\mbox{with}\;\;\;A^{(r)}_j = \frac{\mu}{2^{r+1} \theta^{2^{r+1}-1}}
\prod_{{k=1 \atop k \neq j}}^{2^r-1}
\frac{1}{\xi_k^{(r)}-\xi_j^{(r)}}.
\end{equation}
After performing the integrals, one obtains:
\begin{equation}
\overline{M}^{(r)}_n = 
\sum_{j=1}^{2^r-1} A^{(r)}_j
[\xi_j^{(r)} ]^{-n-1} 
\sum_{m=0}^{\min\{n,2^r\}} { 2^r \choose m}
(1-\alpha \theta^2)^{2^r - m}
\left[\theta^2 \xi_j^{(r)}
\right]^{m} \label{ems}
\end{equation}
This expression shows that the $\overline{M}^{(r)}_n$ may only
{\em approximate} the  ${M}^{(r)}_n$ numbers in a certain limit.
This is seen from the fact that while one must have 
${M}^{(r)}_n = 0$ for $n < 2^r$, and ${M}^{(r)}_{2^r} = 1$, this
is not respected by (\ref{ems}) (it would only be respected if
$\alpha = \theta^{-2}$, however, this is not the case, and the
reason behind this discrepancy is the neglected nonlinearity
from the calculations). The limit, in which the approximation
becomes good is for $r$ large (it means $r \geq 4$) and $n \gg 2^r$.
In this case the sum over $m$ can be performed, and one obtains:
\begin{equation}
\overline{M}^{(r)}_n = 
\sum_{j=1}^{2^r-1} A^{(r)}_j
[\xi_j^{(r)} ]^{-n-1} 
\left[ 1+\theta^2 (\xi_j^{(r)}-\alpha)
\right]^{2^r} \label{ems1}
\end{equation}
The $A^{(r)}_j$ numbers can be calculated in exactly the same way 
we did in the previous section. This leads to:
\begin{equation}
A^{(r)}_j = (-1)^{j+1} \frac{\mu (\xi_j^{(r)} -\alpha)}
{2^r \theta \left[ 1+\theta^2 (\xi_j^{(r)}-\alpha)
\right]^{2^r-1}}\;.
\end{equation}
Inserting it into (\ref{ems1}) it yields:
\begin{equation}
\overline{M}^{(r)}_n = \frac{\mu}{2^r \theta}
\sum_{j=1}^{2^r-1} (-1)^{j+1} 
\frac{\left[ 1+\theta^2 \left(\xi_j^{(r)}-\alpha \right)
\right] \left(\xi_j^{(r)} -\alpha \right)}
{ [\xi_j^{(r)}]^{n+1}} \label{mbar}
\end{equation}
As a check to the correctness of (\ref{mbar}) we can take 
$\mu = 1, \theta = 2$ and $\alpha=1/4$ from the unlabeled case, to 
obtain (\ref{explicit}). 
Equation (\ref{mbar}) explicitely shows the contribution
of each singularity. However, if we want to approximate the
$M^{(r)}_n$ numbers, we should also account for the condition 
$\xi^{(r)}_j < \sqrt{\alpha}$. Using the expression (\ref{sing2}),
this leads to $j < J_r$, where:
\begin{equation}
J_r =  \frac{2^{r+1}}{\pi} 
\mbox{arctg} (\theta \sqrt{\sqrt{\alpha}-\alpha})
\simeq (0.1838035250..) \cdot 2^{r+1}
\end{equation}
Thus, using again (\ref{sing2}):
\begin{equation}
\overline{M}^{(r)}_n = \frac{\mu}{2^r \theta^3}
\sum_{j=1}^{[J_r]} (-1)^{j+1} 
\frac{\tan^2\!{\left(\frac{j\pi}{2^{r+1}}\right)}
\left[ 1 + 
\tan^2\!{\left(\frac{j\pi}{2^{r+1}}\right)}\right]
}
{\left[ \alpha + \theta^{-2} 
\tan^2\!{\left(\frac{j\pi}{2^{r+1}}\right)}\right]^{n+1}}
\label{mbart}
\end{equation}
When the asymptotic limit is generated by the innermost root
$\alpha_r \simeq \xi^{(r)}_1$, i.e., by the first term in
(\ref{mbart}),
one obtains for the topologically self similar ambilateral trees, the scaling behaviour:
\begin{equation}
\overline{M}^{(r)}_n \sim \frac{2\mu\pi^2d^3}{\alpha \theta^3}
e^{-\frac{\pi^2 d^2}{\alpha \theta^2}} n^{-3/2} \gamma^n \label{mbas}
\end{equation}
and therefore
${\cal B} = \gamma = 1/\alpha = 2.4832535..$.

Let us now see how well formula (\ref{mbart}) approximates the
${M}^{(r)}_n $ numbers. To do this, we shall define  the error
$ Q^{(r)}_n = \left[|\overline{M}^{(r)}_n - M^{(r)}_n |
 / M^{(r)}_n \right]
\cdot 100\% $.
For example from the Table in Fig. \ref{tbl1}, $M^{4}_{32} = 
413083691$. The formula above gives 
$[\overline{M}^{(4)}_{32}]= 445781858$, and thus $Q^{(4)}_{32} = 
7.915...\%$. Further error values: $Q^{(5)}_{100}=5.34132...\%$,
$Q^{(5)}_{800}=0.05391...\%$, $Q^{(6)}_{800}=0.003551...\%$.

\section{Conclusions and outlook}

 Combinatorial enumeration of trees is typically difficult to
solve when the set under enumeration obeys symmetry-exclusion
principles, such as for the ambilateral case treated here.
These symmetry-based constraints may arrise in realistic situations
and thus forces us to enumerate {\em classes} of subsets of trees.
In the ambilateral case a class is defined as being formed by
those binary trees that have the same number of leaves and HS
index at the root and can be obtained one from another
via successive reflections with respect to the nodes of the tree.
Certainly, the symmetry operation defining the class must be
an invariant transformation of the topological index (HS in our case).
An other example of such symmetry-operation-generated class-enumeration
is the case of the ``leftist trees'' playing an important
role in the representation of {\em priority queues}, first shown  by
Crane \cite{Crane}, followed by Knuth \cite{Knuth}, who gives
their explicit definition. An elegant enumeration for the
leftist trees, using generating
function formalism was only given very recently
by Nogueira \cite{Nogueira}.

The existing solutions to such class-enumerations on trees (such
as ours and that of Flajolet et. al. \cite{Flajolet} and of
Nogueira \cite{Nogueira}) are obtained via methods taylored for
the particularities of the set and symmetry operation in question.
It is desirable to have, however, at least on a formal level,
a general encompassing theory of class-enumerations of topological
indices. In this direction, powerful methods such as that
of the antilexicographic order method developed by Erd\H{o}s and
Sz\'ekely \cite{ESZ}, or the method of bijection to Schr\"oder trees
developed by Chen \cite{Chen} may turn to be effective after
a suitable extension to include topological indices such as the
Horthon-Strahler index. This, however, stands as an open problem.

\section*{Acknowledgements}

I am especially thankful to Eli Ben-Naim for introducing this
problem to me, and for the many
constructive suggestions while I was working on it.
Useful discussions and comments from
I. Benczik,
T. Brown,
W. Y. C. Chen,
P. L. Erd\H{o}s,
M. Hastings,
G. Istrate and
R. Mainieri
are also gratefully acknowledged. This
work was supported by the Department of Energy 
under contract W-7405-ENG-36.

\end{document}